\documentclass[aps,twocolumn,showpacs,superscriptaddress,groupedaddress,amsmath,amssymb]{revtex4-1}  % for review and submission
\usepackage{graphicx}  % needed for figures
\usepackage{epstopdf}
\usepackage[caption=false]{subfig}
\DeclareMathOperator{\Tr}{Tr}
\def\ket #1{\vert #1\rangle}
\def\bra #1{\langle #1\vert}
\newcommand{\ketbra}[2]{\ensuremath{\ket{#1}\!\bra{#2}}}

\usepackage{longtable}
\usepackage{bbold}
\usepackage{hyperref}
\usepackage{url}
\hypersetup{colorlinks,breaklinks,
citecolor=[rgb]{0.0,0.5,0.5},
    urlcolor=[rgb]{0.0,0.5,0.5},
    linkcolor=[rgb]{0.0,0.5,0.5}}

\begin{document}

\title{Small Codes for Magic State Distillation}
\author{Mark Howard}
\email{m.howard@sheffield.ac.uk}
\affiliation{Department of Physics and Astronomy, University of Sheffield,
Sheffield, UK
}
\affiliation{Institute for Quantum Computing and Department of Applied Mathematics,
University of Waterloo, Waterloo, Ontario, Canada, N2L 3G1
}
\author{Hillary Dawkins}
\affiliation{Institute for Quantum Computing and Department of Physics and Astronomy,
University of Waterloo, Waterloo, Ontario, Canada, N2L 3G1
}

\begin{abstract}
Magic state distillation is a critical component in leading proposals for  fault-tolerant quantum computation. Relatively little is known, however, about how to construct a magic state distillation routine or, more specifically, which stabilizer codes are suitable for the task. While transversality of a non-Clifford gate within a code often leads to efficient distillation routines, it appears to not be a necessary condition. Here we have examined a number of small stabilizer codes and highlight a handful of which displaying interesting, albeit inefficient, distillation behaviour. Many of these distill noisy states right up to the boundary of the known undististillable region, while some distill toward non-stabilizer states that have not previously been considered.
\end{abstract}

%\pacs{}
\maketitle

\section{Introduction}\label{sec:Introduction}
Most efforts towards building a large-scale quantum computer use error-correcting codes to protect the quantum information. However, no matter what code is chosen, the set of gates that are transversal (i.e. manifestly fault-tolerant) will be non-universal \cite{Eastin:2009}, meaning these operations are insufficient for useful quantum computation. Therefore, some additional resource will be required to supplement the limited set of operations. The most promising technique for circumventing this issue is to supplement the non-universal gate set with a supply of special resource states, known as magic states. Having access to pure magic states enables implementation of an additional unitary operation, which ultimately provides us with a universal gate set. The magic state distillation protocol, as introduced by Knill \cite{Knill:2005} and Bravyi and Kitaev \cite{BK}, provides a method of producing these resource states through an iterative procedure in which less pure magic states are consumed to produce a higher purity magic state using only stabilizer operations. 

Here we use small stabilizer codes, which means that the number of qubits, the number of measurements and the number of nonlocal operations involved in each round of distillation are all small. While these are attractive features, it seems that using a small code limits the amount of purification that can occur per round of distillation. For practical purposes, one typically quantifies how many raw magic states are required to produce a single magic state of very high purity (e.g., infidelity of no more than $10^{-10}$). Applying this metric one typically finds that small codes fare much worse than larger codes like the $15$-qubit Reed-Muller code \cite{BK}. If instead we rank codes by their range of applicability, i.e., the threshold noise rate above which purification no longer occurs, then smaller codes seem to fare at least as well as larger codes. Although this may not be a pressing issue experimentally, since it seems reasonable to expect raw state preparation infidelity on the order of $10^{-2}$--$10^{-4}$, it is of foundational interest. Bravyi and Kitaev highlighted the fact that tight magic state distillation routines imply a sharp transition, as noise is decreased, from a circuit that is classically efficiently simulable to one that enables universal quantum computation. The question of tight distillability for qutrit states was addressed in \cite{MH:2015,CAB1}.

Efficient codes for magic state distillation typically exhibit quadratic ($p \mapsto \mathcal{O}(p^2)$) or cubic ($p \mapsto \mathcal{O}(p^3)$) suppression of the error parameter $p$. The codes listed here all exhibit linear error suppression and consequently are not competitive with existing routines in terms of efficiency. Nevertheless we feel that exploring the landscape of codes that achieve distillation is still worthwhile. Reichardt \cite{Reichardt:2009} has previously summarized a handful of known distillation routines exhibiting the best thresholds. 

\begin{figure}[h!]
\includegraphics[scale=0.5]{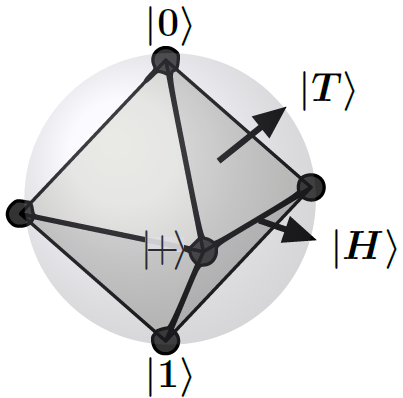}
\caption{\label{fig:octahedron} The stabilizer octahedron inscribed within the Bloch sphere: Six Pauli eigenstates form the vertices of an octahedron. States within the octahedron are provably undistillable so the best one can hope for is to distill states up to the boundary. The two pure non-stabilizer states singled out here, $\ket{H}$ and $\ket{T}$, were shown to be distillable by Bravyi and Kitaev \cite{BK}.}
\end{figure}

A number of the codes presented below achieve tight distillation right up to the boundary of the stabilizer octahedron -- the convex hull of Pauli eigenstates depicted in Fig.~\ref{fig:octahedron} -- whose interior contains states that are provably undistillable. We also present codes that converge, upon iteration, to states that are not those that are typically considered i.e., the $H$-type and $T$-type magic states. None of our codes require twirling (which diagonalizes the state in the $\{\ket{H},\ket{H^\perp}\}$ or $\{\ket{T},\ket{T^\perp}\}$ basis) between rounds.

\section{Background}\label{sec:Background}

Here we briefly summarize notation and terminology associated with both magic state distillation and with the type of quantum codes that we have used.

\subsection{Magic state distillation, thresholds and yields}\label{sec:Magic_state_distillation_thresholds_and_yields}
Magic state distillation routines are described in terms of stabilizer error correcting codes, which in turn are described by a set of generators $\{G_i\}$ consisting of Pauli operators. The protocol (see e.g. \cite{CB2}) consists of iteratively applying the following steps: 
\begin{enumerate}
\item Prepare $n$ copies of the input state $\rho_{in}^{\otimes n}$, where $n$ is the size of the code
\item Perform Pauli measurements corresponding to each of the $n-k$ generators $G_i$, and postselect on the desired outcome 
\item Optionally perform a Clifford transformation based on the measurement outcome. 
\end{enumerate}
When successful, the output state(s) will be purified in the direction of the target magic state. Typically $k=1$ and consequently an $[[n,k=1,\text{distance}]]$ stabilizer code is used (see \cite{Meier:2013, Bravyi:2012, Jones:2013} for $k\geq 2$ however). 

For simplicity, one can assume that input states suffer from depolarizing noise and take the form
\begin{align}
%\rho_{M} = (1-p)\ketbra{M}{M} +p\frac{1\!\!1_3}{d}~, \label{eqn:depol}
\rho_{M} = (1-p)\ketbra{M}{M} +p\frac{\mathbb{1}_2}{2}~, \label{eqn:depol}
\end{align}
where $\ket{M}$ is the target magic state. The maximum value of $p$ for which distillation is possible even in principle, which we call $p_{oct}$, is when $\rho_M$ crosses the boundary into the stabilizer octahedron. With reference to Fig~\ref{fig:octahedron}, $\ket{H}$-type states are of the form $\ket{H}=(
\ket{0}+e^{i \pi/4}\ket{1})/\sqrt{2}$ or any image of this under a Clifford gate, while $\ket{T}$-type states have Bloch vectors of the form $(\pm1, \pm1, \pm1)/\sqrt{3}$. The simple geometry allows us to find $p_{oct} = (1-1/\sqrt{2}) \approx 0.2929$ for $\ket{H}$-type states and $p_{oct} = (1-1/\sqrt{3}) \approx 0.4226$ for $T$-type states.   

The yield \cite{CAB3} of a magic state distillation routine is defined as
\begin{align}
Y(p,p_{out}) = \prod_{k = 1...N}\frac{p_s^{(k)}}{n} \label{eqn:yield}
\end{align}
where $N$ is the total number of iterations needed to obtain $p_{out}$ starting at initial error rate $p$, and $p_s^{(k)}$ is the probability of success on the $k^{th}$ iteration. This quantity relates to the efficiency/resource overhead of a magic state distillation routine.

\subsection{Codeword stabilized quantum codes}\label{subsec:Codeword_stabilized_quantum_codes}

The Pauli measurements associated with a stabilizer code can be written in binary symplectic notation so that $G=(X^{x_1}\otimes X^{x_2}\otimes \ldots \otimes X^{x_n})(Z^{z_1}\otimes Z^{z_2}\otimes \ldots \otimes Z^{z_n})$ becomes $(x_1,x_2,\ldots,x_n|z_1,z_2,\ldots,z_n)$. An $n$-qubit stabilizer code with $n$ generators has $k=0$ and consequently specifies a 1-dimensional subspace of Hilbert space i.e, a stabilizer state. By applying local Cliffords (the unitaries that map Pauli operators to Pauli operators under conjugation), a generic stabilizer state $(X|Z)$ can be brought to the form $(\mathbb{1}_n|\Gamma)$ where $\Gamma$ is the adjacency matrix of the graph. When stabilizer states take this particular form they are called graph states, sometimes denoted $\ket{\Gamma}$.

\begin{figure}%
\centering
\subfloat[Graph with labelled vertices]{\label{fig:GraphCircuit_graph}\includegraphics[scale=0.5]{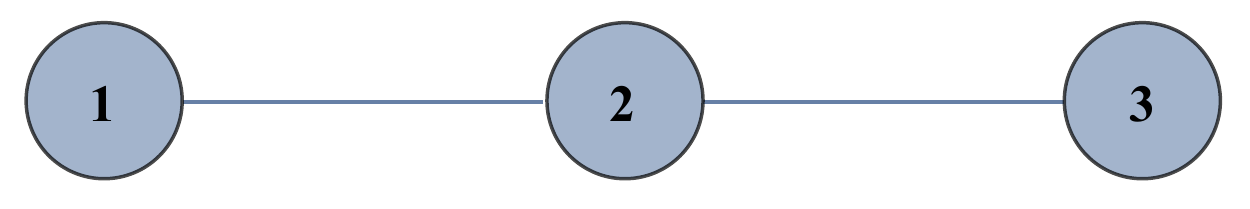}}\\
\subfloat[The graph state preparation circuit for the above graph]{\label{fig:GraphCircuit}\includegraphics[scale=1]{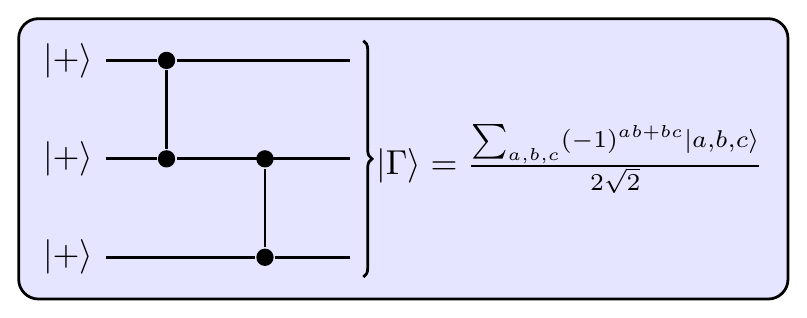}}
\caption{Creating the graph state associated with a given graph amounts to performing controlled-$Z$ rotations between qubits whose vertices are connected by an edge in the graph.}
\label{fig:example}
\end{figure}

The majority of magic state distillation routines presented here will be described by codeword stabilized (CWS) codes (see \cite{Cross:2009}). In this representation, the code is given by a graph state written as an adjacency matrix, $\Gamma \in \mathbb{Z}_2^{n\times n}$, as well as a classical codeword, $w \in \mathbb{Z}_2^n$. We may choose the logical basis state $\ket{0_L}$ of our code to be the state 
\begin{align}
\ket{0_L} = \ket{\Gamma} = \sum_{x\in \mathbb{Z}_2^n}i^{x^T\Gamma x} \ket{x}. \label{eqn:graphket}
\end{align} 
Subsequently the logical $\ket{1_L}$ operator is given by
\begin{align}
\ket{1_L} = Z^{w}\ket{0_L} = \sum_{x\in \mathbb{Z}_2^n} i^{x^T\Gamma x + 2w^Tx }\ket{x}, \label{eqn:1ket}
\end{align} 
where $w$ is the codeword and $Z^{w}$ denotes the product of $Z$ operators
\begin{align}
Z^{w} = Z^{w_1}\otimes ... \otimes Z^{w_n}. \label{eqn:Zops}
\end{align} 
The unitary encoding/decoding associated with these codes is quite straightforward as illustrated in Fig.~\ref{fig:GraphCircuit} -- it involves controlled-$Z$ rotations applied to qubits whose vertices are connected by edges in the graph.
A convenient way of visualizing CWS codes is to display the graph and highlight the subset of vertices $1\leq i \leq n$ such that $w_i=1$ see Fig.~\ref{fig:GraphsA},\ref{fig:GraphsB} and \ref{fig:nontightdistillers}.

\section{Results}\label{sec:Results}
All stabilizer codes are local Clifford equivalent to some CWS stabilizer code. Consequently CWS stabilizer codes form a subset of all stabilizer codes and so checking all combinations of graphs, $\Gamma$, and codewords, $w$, may still miss stabilizer codes that are useful for distillation. Nevertheless, for $n\leq 4$ qubits we iterated over all graphs on $n$ vertices and for $n\in\{5,6\}$ qubits we iterated over non-isomporphic graphs on $n$ vertices. We were primarily focused on recording those CWS codes that achieve tight distillation but we also noted a number of non-tight codes that distilled to target states other than $\ket{H}$.

With the exception of one code, the codes that we present will be depicted graphically. This concise representation is possible because of the CWS formalism that we described in \ref{subsec:Codeword_stabilized_quantum_codes}. All the codes that we present converge, upon repeated iteration, to a pure state on the surface of the Bloch sphere. We give this point in coordinates $(x,y,z)$ where $x=\Tr(\ketbra{M}{M} X)$ etc. Codes that we describe below as being tight obey the following property: all non-stabilizer states in the same quadrant as $\ket{M}$ are distillable. For instance, if $\ket{M}$ has Bloch vector $(x,0,z)$ with $x,z >0$ then all states $\rho$ satisfying $\Tr(X \rho)+\Tr(Z\rho) >1$ converge to $\ket{M}$ under repeated iteration. (We confirmed this numerically by taking a random sample of 1000 points from the relevant region).

\subsection{Codes achieving tight distillation}\label{subsec:Codes_achieving_tight_distillation}

\begin{figure}[h!]
\centering
\subfloat[This code converges to $(x,y,z)=(\sin \theta,0,\cos \theta)$ where $\theta=\arctan \sqrt{(\sqrt{5}-1)/2}$]{\label{fig:Graphs-a}\includegraphics[scale=0.5]{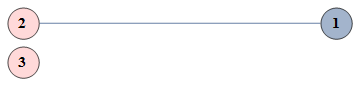}}\\
\subfloat[This code converges to $(x,y,z)=(0.66796,0.,0.7442)$]{\label{fig:Graphs-b}\includegraphics[scale=0.5]{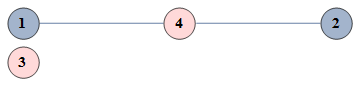}}\\
\subfloat[This code converges to $(x,y,z)=(0.81281,0,0.58252)$]{\label{fig:Graphs-b}\includegraphics[scale=0.5]{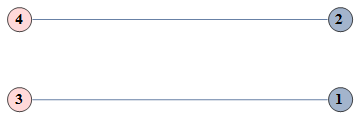}}\\
\subfloat[This code converges to $(x,y,z)=(0.64969,0,0.7602)$]{\label{fig:Graphs-c}\includegraphics[scale=0.35]{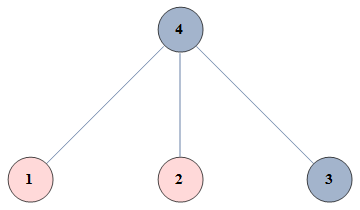}}\\
\subfloat[This code converges to $(x,y,z)=(1,0,1)/\sqrt{2}$]{\includegraphics[scale=0.5]{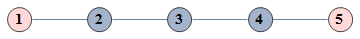}}
\caption{Codeword stabilized quantum codes with tight distillation thresholds. The classical codeword $w$ associated with each graph is the binary vector with a ``$1$'' in lighter pink positions and ``$0$'' in darker blue positions.}
\label{fig:GraphsA}
\end{figure}

\begin{figure}[h!]
\centering
\subfloat[This code converges to $(x,y,z)=(0.84893,0.,0.52851)$]{\includegraphics[scale=0.3]{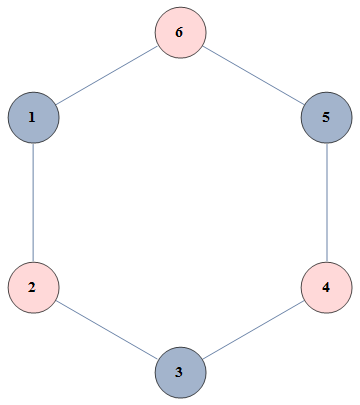}}\\
\subfloat[This code converges to $(x,y,z)=(0.63544,0.,0.77215)$]{\includegraphics[scale=0.35]{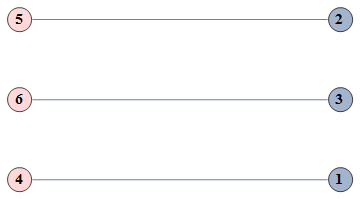}}\\
\subfloat[This code converges to $(x,y,z)=(0.81281,0.,0.58252)$]{\includegraphics[scale=0.35]{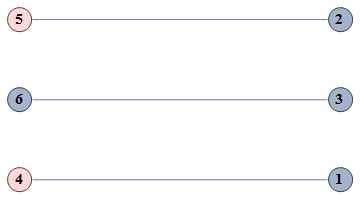}}\\
\subfloat[This code converges to $(x,y,z)=(0.84534,0.,0.53423)$]{\includegraphics[scale=0.5]{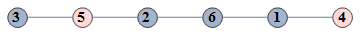}}\\
\subfloat[This code converges to $(x,y,z)=(0.58252,0.,0.81281)$]{\includegraphics[scale=0.5]{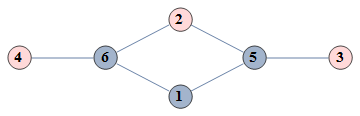}}
\caption{Codeword stabilized quantum codes with tight distillation thresholds. The classical codeword $w$ associated with each graph is the binary vector with a ``$1$'' in lighter pink positions and ``$0$'' in darker blue positions.}
\label{fig:GraphsB}
\end{figure}

\begin{figure}[h]
\includegraphics[scale=1.4]{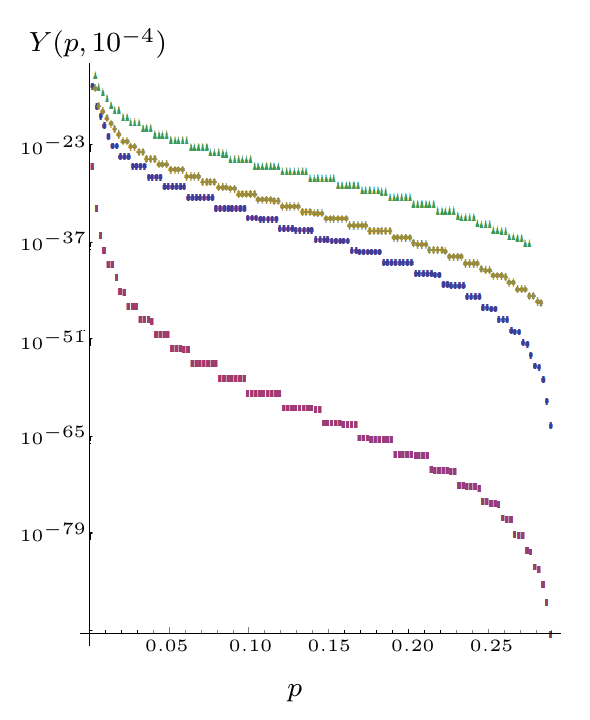}
\caption{\label{fig:yield} Curves showing the yield (efficiency) of various tight distillation routines as a function of input noise rate $p$. From bottom to top we have $(i)$ the 7-qubit Steane code as applied by Reichardt \cite{Reichardt} $(ii)$ the 5-qubit code in Fig.~\ref{fig:GraphsA}(e),  $(iii)$ 3-qubit code from Fig.~\ref{fig:GraphsA}(a) and $(iv)$ the 3-qubit code in Eq.~\eqref{eqn:3aGen}.}
\end{figure}

A 3-qubit code with generators 
\begin{align}
\begin{matrix} G_1 & Z & I & Z \\ G_2 & X & Z & X \\ Z_L & X & X & Y \\ X_L & I & X & Z \\ \end{matrix} \label{eqn:3aGen}
\end{align}
corresponding to the logical basis states 
\begin{align}
\ket{0_L} = \frac{1}{2}\begin{pmatrix}1\\0\\-i\\0\\0\\1\\0\\i\\ \end{pmatrix},  \ket{1_L} = \frac{1}{2}\begin{pmatrix}-i\\0\\1\\0\\0\\-i\\0\\-1\\ \end{pmatrix} \label{eqn:3alogops}
\end{align}
distills an equatorial state in the $y$-$z$ plane with Bloch coordinates $(0,-.83929, -.54369)$ up to a tight error threshold of $p_{oct} = .276921$.

Other codes achieving tight distillation are given in pictorial form in Fig.~\ref{fig:GraphsA} and Fig.~\ref{fig:GraphsB}. It is a straightforward exercise to recover the generators and logical operators if necessary. The 5-qubit code in  Fig.~\ref{fig:GraphsA}(e) distills to the $\ket{H}$-type magic state but is more efficient than the (also tight) 7-qubit Steane code. The efficiency of these codes is compared in Fig.~\ref{fig:yield}.

\subsection{Codes not achieving tight distillation}\label{subsec:Codes_not_achieving_tight_distillation}

Here we highlight a small number of codes that, even though they do not achieve tight distillation, we still find to be noteworthy.

\begin{figure}[h!]
\centering
\subfloat[This is the well-known $5$-qubit code in CWS form, which Bravyi and Kitaev showed distilled $
\ket{T}$]{\label{fig:Graphs-a}\includegraphics[scale=0.35]{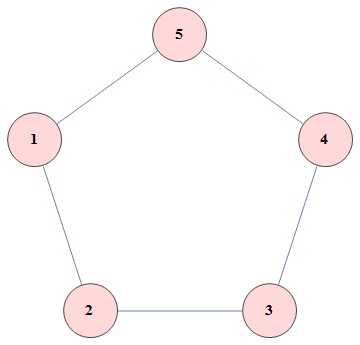}}\\
\subfloat[This code also distills to $\ket{T}$ albeit less efficiently and with a worse threshold than the above code.  Note that a trivial $Z$ correction must be applied in between rounds]{\label{fig:Graphs-b}\includegraphics[scale=0.5]{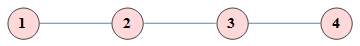}}\\
\subfloat[This code converges to a state Clifford-equivalent to $\ket{\pi/3}=(\ket{0}+e^{i \pi/3}\ket{1})/\sqrt{2}$. The limiting state has Bloch vector  $(x,y,z)=(\frac{1}{2},0,\frac{\sqrt{3}}{2})$]{\label{fig:Graphs-b}\includegraphics[scale=0.5]{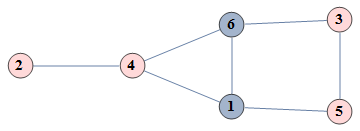}}\\
\subfloat[This code converges to $(x,y,z)=(0.60965,0.,0.79267)$]{\includegraphics[scale=0.4]{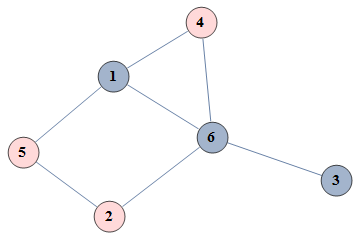}}
\caption{Codeword stabilized quantum codes that do not achieve tight thresholds}
\label{fig:nontightdistillers}
\end{figure}

The $\ket{T}$-type states (depicted in Fig.~\ref{fig:octahedron}) were shown to be distillable by Bravyi and Kitaev \cite{BK} using the perfect $[[5,1,3]]$ code. We are not aware of any additional routines for $\ket{T}$ states that have subsequently been developed. This is in marked contrast to $\ket{H}$-type distillation for which a number of codes have been found. For this reason we note the existence of a $4$-qubit code in Fig.~\ref{fig:nontightdistillers}(b) that also converges to $\ket{T}$-type states. A visual comparison with the $[[5,1,3]]$ code in Fig.~\ref{fig:nontightdistillers}(a) indicates that they are closely related. Unfortunately, the threshold for the new code is worse than that of the $[[5,1,3]]$ code. Finding tight distillation routines for the $\ket{T}$ direction was already known to be more complex than the $\ket{H}$ case because of a no-go theorem in \cite{CB1}.

In  Fig.~\ref{fig:nontightdistillers}(c) we depict a code that distills $\ket{\pi/3}=(\ket{0}+e^{i\pi/3}\ket{1}){/\sqrt{2}}$. This is particulary interesting as the associated gate $U=\textrm{diag}(1,e^{i \pi/3})$ is not transversal for any stabilizer code \cite{Anderson:2014} which prevents it from being distillable by the most commonly used distillation techniques.

It is of interest to find distillation routines that distill $\ket{V}$-type magic states, which look like $(x,y,z)=(\frac{3}{5},0,\frac{4}{5})$ in the Bloch sphere picture. Supplementing Cliffords with such states leads to a set of gates -- the $V$-basis -- that is highly efficient for gate synthesis \cite{Bocharov:2013}. We did not find such a code but in  Fig.~\ref{fig:nontightdistillers}(d) we depict a code that converges to a nearby state $(x,y,z)=(0.60965,0.,0.79267)$.

\section{Conclusions}\label{sec:Conclusions}
We have presented a collection of qubit magic state distillation schemes using small stabilizer codes, most of which achieve tight distillation up to the edge of the stabilizer octahedron. Two codes that are not tight were still noted because they converge to $\ket{T}$ and $\ket{\pi/3}$ respectively. Our distillation routines converge to equatorial states, which means that they can be used to implement rotations about a Pauli axis using standard half-teleportation techniques \cite{BK}. We have noted that inefficiency of our routines relates to linear error suppression and leave as open question whether there is some way of boosting their efficiency by combining them with other codes or techniques. Another possible avenue is to use these codes to convert non-stabilizer states of one type to another, as was done in e.g.,  \cite{GDC:2013}. For example, if we have access to almost pure $\ket{H}$ states  (after using the Reed-Muller code, say) then input these to the $\ket{\pi/3}$ routine, the output is an almost pure state somewhere between $\ket{H}$ and $\ket{\pi/3}$. Further analysis of the relative merits of methods such as this is left for future work.

\clearpage
\section{Acknowledgments}
The authors acknowledge financial support from the Government of Canada through NSERC through the CGS-M program and the discovery grant program, as well as the U.S. Army Research Office through Grant No. W911NF-14-1-0103, and FQXI. MH was supported by EPSRC (grant EP/M024261/1).

\end{document}